\documentclass{acm_proc_article-sp}

\begin{document}

\title{Role of Weak Ties in Link Prediction of Complex Networks}
%
%
%
%
%

\numberofauthors{2} 
%
\author{
\alignauthor Linyuan L\"{u}\\
       \affaddr{Department of Physics, University of Fribourg}\\
       \affaddr{Fribourg 1700, Switzerland}\\
       \email{babyann519@hotmail.com}
\alignauthor Tao Zhou\\
       \affaddr{Joint Laboratory of Complex Systems \& Information Physics, University of Science and Technology of China and University of Fribourg}\\
       \email{zhutou@ustc.edu}
}

\maketitle
\begin{abstract}
Plenty of algorithms for link prediction have been proposed and were
applied to various real networks. Among these works, the weights of
links are rarely taken into account. In this paper, we use local
similarity indices to estimate the likelihood of the existence of
links in weighted networks, including \emph{Common Neighbor},
\emph{Adamic-Adar Index}, \emph{Resource Allocation Index}, and
their weighted versions. In both the unweighted and weighted cases,
the resource allocation index performs the best. To our surprise,
the weighted indices perform worse, which reminds us of the
well-known \emph{Weak Tie Theory}. Further experimental study shows
that the weak ties play a significant role in the link prediction
problem, and to emphasize the contribution of weak ties can
remarkably enhance the predicting accuracy.
\end{abstract}




\section{Introduction}
Many complex systems can be well described by networks where nodes
present individuals or agents, and links denote the relations or
interactions between nodes. Recently, the link prediction of complex
networks has attracted more and more attention from computer
scientists \cite{Getoor} and physicists \cite{newman2,Redner2008}.
Link prediction aims at estimating the likelihood of the existence
of a link between two nodes, based on the observed links and the
attributes of the nodes. For example, classical information
retrieval can be viewed as predicting missing links between words
and documents \cite{salton}, and the process of recommending items
to a user can be considered as a link prediction problem in the
user-item bipartite network \cite{Zhou2007}.

The problem of link prediction can be categorized into two classes:
One is the prediction of existed yet unknown links for sampling
networks, such as food webs, protein-protein interaction networks
and metabolic networks; the other is the prediction of links that
may exist in the future of evolving networks, like on-line social
networks. In addition, the link prediction algorithms can also be
used to generate some artificial links to help the further network
analysis, such as the classification problem in partially labeled
networks \cite{Macskassy2007,Gallagher2008}. Some algorithms based
on Markov chains \cite{Sarukkai,Zhu,Bilgic} and machine learning
\cite{Popescul,Wang} have been proposed recently, and another group
of algorithms are based on the definition of node similarity. In
this paper, we concentrate on the latter. Node similarity can be
defined by using the essential attributes of nodes, namely two nodes
are considered to be more similar if they have many common features.
However, the essential features of nodes are usually not available,
and thus the mainstream of similarity-based link prediction
algorithms consider only the observed network structure.
Liben-Nowell and Kleinberg \cite{Liben-Nowell} systematically
compared some structure-based node similarity indices for link
prediction problem in co-authorship networks, and Zhou \emph{et al.}
\cite{Linyuan,Lu2009} studied nine well-known local similarity
indices on six real networks extracted from disparate field, as well
as proposed two new local indices.

Up to now, most studies of link prediction do not take weights of
links into consideration. Murata \emph{et al.} \cite{Murata}
proposed three weighted similarity indices, as variants of Common
Neighbors, Adamic-Adar and Preferential Attachment indices
respectively. They applied these indices to the networks of
\emph{Question-Answer Bulletin Boards System}, and the results show
that with the consideration of weights the prediction accuracy can
be enhanced. To our surprise, when we apply the weighted indices to
co-authorship networks and the US air transportation network, we
find that the weighted indices perform even worse than the
unweighted ones. Actually, Liben-Nowell and Kleinberg
\cite{Liben-Nowell} reported the similar observation for weighted
Katz index. These unexpected results remind us of the well-known
\emph{Weak Tie Theory} \cite{Granovetter}. Further experimental
study shows that the weak ties play a significant role in the link
prediction problem, and to emphasize the contribution of weak ties
can remarkably enhance the predicting accuracy.
\begin{table*}
\centering \caption{Algorithmic accuracy, measured by precision.
Each number is obtained by averaging over 100 implementations with
independently random divisions of testing set and probe set. The
numbers inside the brackets denote the standard derivations. For
example, 0.592(48) means the precision is 0.592, and the standard
derivation is 0.048. The abbreviation, WCN*, WAA* and WRA*,
represents the highest precisions obtained by Eqs. (7-9),
respectively. The corresponding optimal values of $\alpha$ are shown
in Table 2.}
\begin{tabular}{cccccccccc} \hline
      &CN&WCN&WCN*&AA&WAA&WAA*&RA&WRA&WRA*\\ \hline
USAir     &0.592(48)&0.443(48)&0.617(45)&0.606(49)&0.517(50)&0.639(48)&0.626(39)&0.558(48)&0.633(41) \\
\hline NetScience&0.822(51)&0.202(37)&0.822(51)&0.957(33)&0.681(43)&0.959(22)&0.962(18)&0.978(14)&0.978(14)\\
\hline
CGScience &0.625(59)&0.299(45)&0.782(57)&0.780(49)&0.292(51)&0.917(37)&0.963(15)&0.938(17)&0.969(16) \\
\hline\end{tabular}
\end{table*}
\section{Data and Method}
Considering an undirected simple network $G(V,E)$, where $V$ is the
set of nodes and $E$ is the set of links. The multiple links and
self-connections are not allowed. For each pair of nodes, $x,y\in
V$, we assign a score, $s_{xy}$, according to a given similarity
measure. Higher score means higher similarity between these two
nodes, and vice versa. Since $G$ is undirected, the score is
supposed to be symmetry, say $s_{xy}=s_{yx}$. All the nonexistent
links are sorted in the decreasing order according to their scores,
and the links in the top are most likely to exist. To test the
algorithmic accuracy, the observed links, $E$, is randomly divided
into two parts: the training set, $E^T$, is treated as known
information, while the probe set, $E^P$, is used for testing and no
information therein is allowed to be used for prediction. Clearly,
$E=E^T\cup E^P$ and $E^T\cap E^P=\varnothing$. In this paper, the
training set always contains 90\% of links, and the remaining 10\%
of links constitute the probe set. To quantify the prediction
accuracy, we use a standard metric called \emph{precision}, which is
defined as the ratio of relevant items selected to the number of
items selected. We focus on the top $L$ predicted links (in this
paper, we set $L=100$), if there are $L_{r}$ relevant links (i.e.,
the links in the probe set), the precision equals $L_{r}/L$.
Clearly, higher precision means higher prediction accuracy.

The empirical data used in this paper include (i) USAir.---the US
air transportation network, which contains 332 airports and 2126
airlines (see \emph{Pajak Datasets}). The weight of a link is the
frequency of flights between two airports. (ii) NetScience.---the
co-authorship network of 1589 scientists who are themselves working
on network science \cite{Newman3}. Here, the weight between two
scientists is not simply the number of papers they co-authorized.
According to \cite{Newman4}, if a paper has $n$ coauthors, then the
weight of each pair of authors contributed by this paper is
$1/(n-1)$. For two scientists, the final weight of their link is
obtained by summing up the weights contributed by all their
co-authorized papers. (iii) CGScience.---the co-authorship network
in computational geometry till February 2002 (see \emph{Pajek
Datasets}). This network contains 7343 authors and 11898 links. Two
authors are linked if they wrote at least a common paper/book. The
weight of a link is assigned by directly counting the number of
common papers/books.
\begin{figure}
\centering \epsfig{file=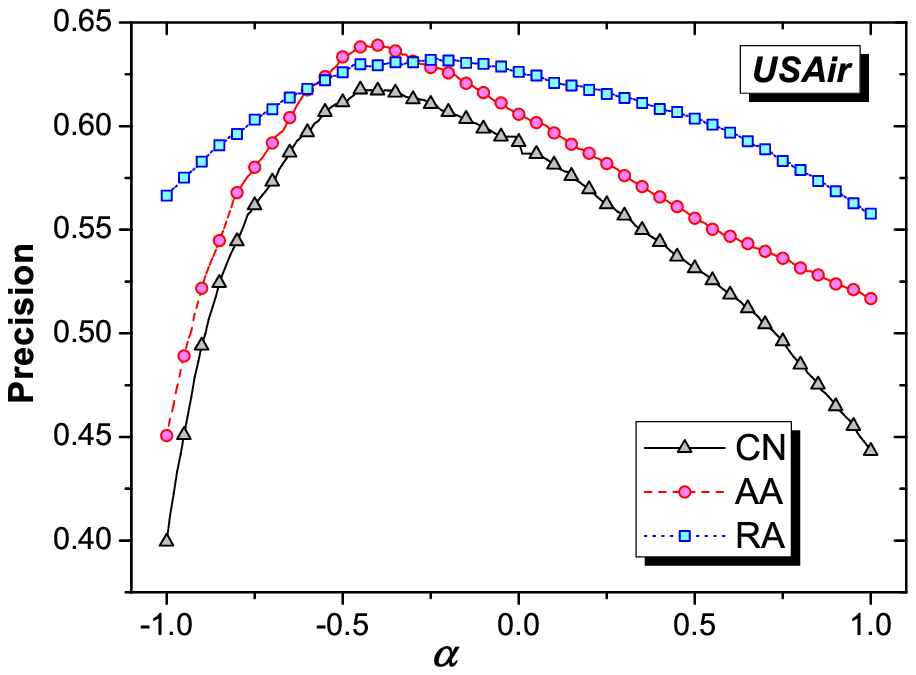, height=2.8in, width=3.5in}
\centering \epsfig{file=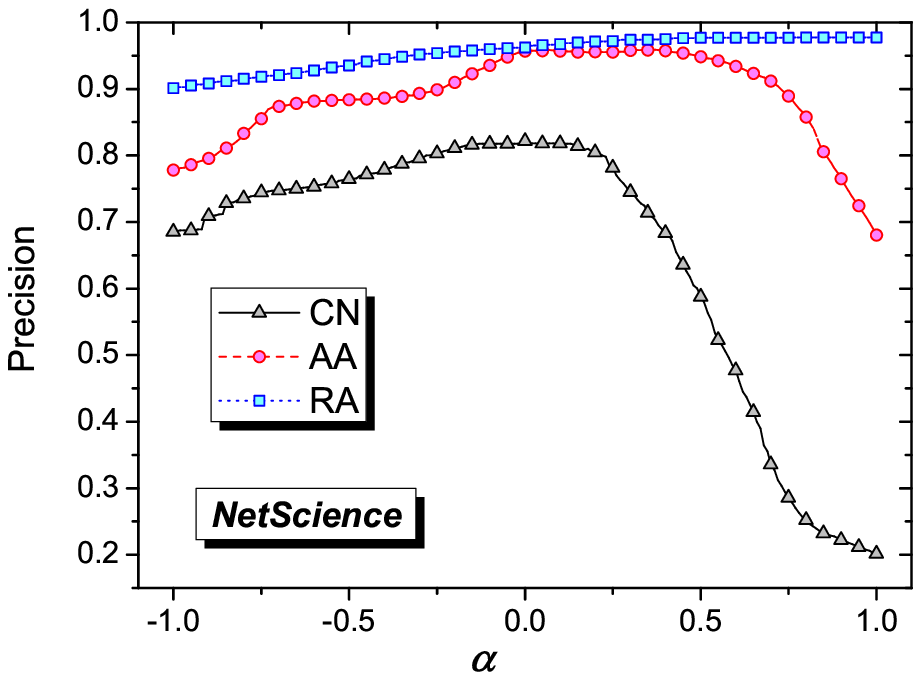, height=2.8in, width=3.5in}
\centering \epsfig{file=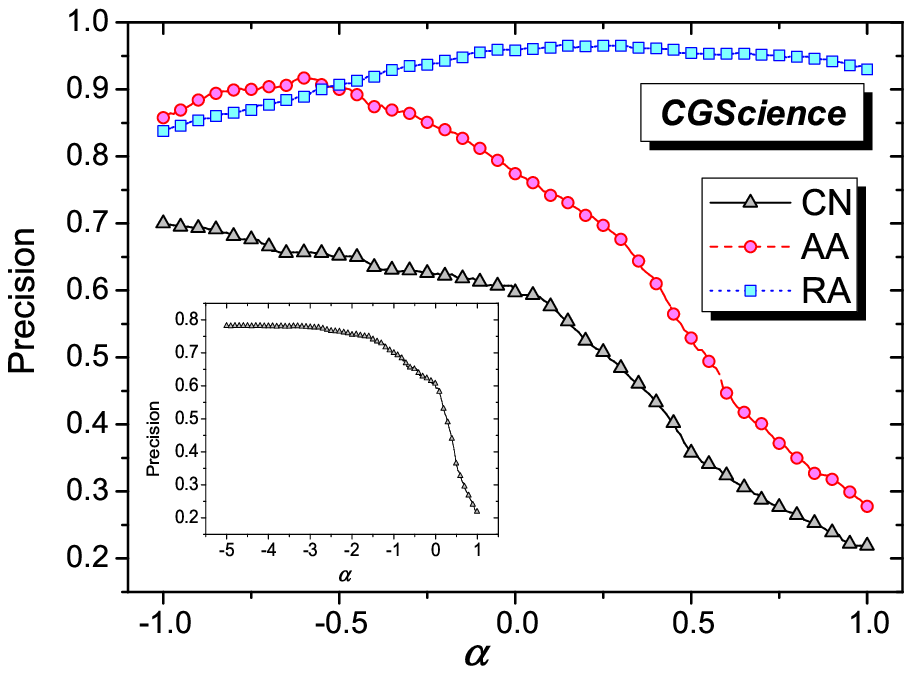, height=2.8in, width=3.5in}
\caption{Precision as a function of $\alpha$ for USAir, NetScience
and CGScience: CN (triangles), AA (circles) and RA (squares). The
inset in the plot for CGScience shows the precision of CN for
$\alpha \in [-5,1]$. Each data point is obtained by averaging over
100 realizations, each of which corresponds to an independent
division of training set and testing set.}
\end{figure}
\section{Unweighted Similarities Based on Local Information}
Among many similarity indices, Liben-Nowell and Kleinberg
\cite{Liben-Nowell} showed that the \emph{Common Neighbors} (CN) and
\emph{Adamic-Adar} (AA) index \cite{Adamic} perform the best, which
has been further demonstrated by systematically comparing CN, AA
index with seven other well-known local similarity indices
\cite{Linyuan}. In addition, Zhou \emph{et al.} \cite{Linyuan}
proposed a new index named \emph{Resource Allocation} (RA) index,
which can beat both CN and AA index. Therefore, in this paper, we
concentrate on CN, AA index and RA index, whose definitions are as
following.

(i) CN. In common sense, two nodes, $x$ and $y$, are more likely to
form a link in the future if they have many common neighbors. Let
$\Gamma(x)$ denote the set of neighbors of node $x$. The simplest
measure of the neighborhood overlap is the directed count:
\begin{equation}
s_{xy}=|\Gamma(x)\cap \Gamma(y)|,
\end{equation}
where $|Q|$ is the cardinality of the set $Q$.

(ii) AA index. It refines the simple counting of common neighbors by
giving the lower-connected neighbors more weights, as:
\begin{equation}
s_{xy}=\sum_{z\in\Gamma(x)\cap \Gamma(y)}\frac{1}{\texttt{log}k(z)},
\end{equation}
where $k(z)$ is the degree of node $z$, namely $k(z)=|\Gamma(z)|$.

(iii) RA index. Considering a pair of nodes, $x$ and $y$, which are
not directly connected. The node $x$ can send some resource to $y$,
with their common neighbors playing the role of transmitters. In the
simplest case, we assume that each transmitter has a unit of
resource, and will equally distribute to all its neighbors. As a
results the amount of resource $y$ received is defined as the
similarity between $x$ and $y$, which is:
\begin{equation}
s_{xy}=\sum_{z\in\Gamma(x)\cap \Gamma(y)}\frac{1}{k(z)}.
\end{equation}

Empirical analysis shows that \cite{Linyuan} comparing with CN and
AA, RA can enhance the prediction accuracy measured by the area
under a receiver operating characteristic curve (AUC)
\cite{Hanley1982}, especially for the networks with large average
degrees (in such cases, the difference between RA and AA is large).
AUC takes into account the whole ranking, while precision only
concentrates on the top $L$ predicted links. As shown in Table 1,
subject to precision, RA still performs remarkably better than CN
and AA. Here comes a simple but significant result, the RA index
outperforms CN and the AA index, and thus can find its applications
in better characterize the proximity of nodes in networks.

\section{Weighted Similarities}
The similarity indices mentioned in the last section only consider
the binary relations among nodes, however, in the real world, links
are naturally weighted, which may represent the transportation load
between two airports in a airline network or the number of
co-authorized papers in a co-authorship network. We expect the
similarity index taking into account link weights can give better
predictions. Murate and Moriyast \cite{Murata} proposed a simple way
to extend a similarity index for binary network to a weighted index.
Following this method, the weighted CN, weighted AA index and
weighted RA index (denoted by WCN, WAA and WRA, respectively) are:
\begin{equation}
s_{xy}=\sum_{z\in\Gamma(x)\cap \Gamma(y)}{w(x,z)+w(z,y)},
\end{equation}
\begin{equation}
s_{xy}=\sum_{z\in\Gamma(x)\cap
\Gamma(y)}\frac{w(x,z)+w(z,y)}{\texttt{log}(1+s(z))},
\end{equation}
\begin{equation}
s_{xy}=\sum_{z\in\Gamma(x)\cap \Gamma(y)}\frac{w(x,z)+w(z,y)}{s(z)}.
\end{equation}
Here, $w(x,y)=w(y,x)$ denotes the weight of link between nodes $x$
and $y$, and $s(x)=\sum_{z\in\Gamma(x)}w(x,z)$ is the strength of
node $x$. Note that, since $s(z)$ may be smaller than 1 we use
$\texttt{log}(1+s(z))$ in Eq. (5) to avoid negative score.

To our surprise, when we apply the weighted indices to the three
experimental networks, as shown in Table 1, we find that except WRA
in NetScience, the weighted indices perform obviously worse than
their corresponding unweighted versions. Especially for CN, with
consideration of the weights the precisions are sharply decreased.
These unexpected results remind us of the well-known \emph{Weak Ties
Theory} \cite{Granovetter}, which states that the people usually
obtain useful information or opportunities through the acquaintances
but not the close friends, namely the weak ties in their friendship
network paly a significance role. Recently, Onnela \emph{et al.}
\cite{Onnela2007} demonstrated that the weak ties mainly maintain
the connectivity in mobile communication networks, and Csermely
found that the weak ties may maintain the stability of biological
systems \cite{Csermely2004}. In contrast, the role of weak ties in
link prediction problem has not been investigated yet.

\section{Role of Weak Ties}
In this section, we provide a start point to investigate the role of
weak ties in link prediction by introducing a free parameter,
$\alpha$, to control the relative contributions of weak ties to the
similarity measure. The parameter-dependent indices for WCN, WAA and
WRA are:
\begin{equation}
s_{xy}=\sum_{z\in\Gamma(x)\cap
\Gamma(y)}{w(x,z)^{\alpha}+w(z,y)^{\alpha}},
\end{equation}
\begin{equation}
s_{xy}=\sum_{z\in\Gamma(x)\cap
\Gamma(y)}\frac{w(x,z)^{\alpha}+w(z,y)^{\alpha}}{\texttt{log}(1+s(z))},
\end{equation}
\begin{equation}
s_{xy}=\sum_{z\in\Gamma(x)\cap
\Gamma(y)}\frac{w(x,z)^{\alpha}+w(z,y)^{\alpha}}{s(z)},
\end{equation}
where $s(x)=\sum_{z\in\Gamma(x)}w(x,z)^{\alpha}$. When $\alpha=0$,
$s(x)$ is the degree of node $x$, and the indices degenerate to the
unweighted cases. When $\alpha=1$, the indices is equivalent to the
simply weighted indices, as shown in Eqs. (4-6). The numerical
results are given in Figure 1, Table 1 and Table 2. For all cases,
the optimal values of $\alpha$ are all smaller than 1. That is to
say, the weak links play more important role in the link prediction
than indicated by their weights. A big surprise is that sometimes
the optimal values of $\alpha$ are negative, that is to say, the
weak links actually play more important role than the strong links.
Although it is well-known that the weak ties mainly maintain the
network connectivity, this result is still striking for us.
\begin{table}
\centering \caption{Optimal values of the parameter $\alpha$ subject
to the highest precisions. For CGScience, with the decreasing of
$\alpha$ the precision increases monotonously and eventually reaches
a stable value, 0.782, at the point $\alpha=-4.15$.}
\begin{tabular}{cccc} \hline
      &WCN*&WAA*&WRA*\\ \hline
USAir     &-0.41&-0.40&-0.24\\ \hline
NetScience&0.00&0.36&0.80\\
\hline
CGScience &-4.15&-0.60&0.13 \\
\hline\end{tabular}
\end{table}

\section{Conclusions and Discussion}
In this paper, we applied three local similarity indices, Common
Neighbor, Adamic-Adar index and Resource Allocation index, to the
link prediction problem of three empirical networks, USAir,
NetScience and CGScience. We found that our previously proposed
index, RA \cite{Linyuan}, performs the best. Furthermore, with the
consideration of weights, we tested three weighted variants of CN,
AA and RA, denoted by WCN, WAA and WRA. To our surprise, the
precision of weighted indices perform even worse than their
corresponding unweighted versions. These unexpected results remind
us the weak ties theory \cite{Granovetter} which claims that the
links with small weights yet play an important role in social
network. Extensive experimental study shows that the weak ties play
a significant role in the link prediction problem, and to emphasize
the contribution of weak ties can remarkably enhance the predicting
accuracy. Sometimes, in the optimal cases, the weak ties contribute
more than the strong ties. In another word, the weak links in such
network are not as weak as their weights suggested.

Although the prediction accuracies of both the unweighted indices
(Eqs. (1-3)) and the simply weighted indices (Eqs. (4-6)) can be
further improved by introducing the parameter $\alpha$ (Eqs. (7-9)),
this paper does not aim at highlighting these parameter-dependent
indices. Instead, we attempt to uncover the role of weak ties in the
link prediction problem. We hope this paper can provide a start
point for the possible \emph{weak ties theory in information
retrieval}.

\bibliographystyle{abbrv}
\bibliography{sigproc}
\end{document}